\documentclass[preprint,aps,pra,showpacs,floatfix]{revtex4}
\usepackage{graphicx}
\usepackage{times}
\usepackage{nicefrac}
\usepackage{amsmath}
\usepackage{amsfonts}
\usepackage{amssymb}
\usepackage{amsthm}
\usepackage{epsf}
\usepackage{bm}
\usepackage{bbm}
\usepackage{times}

\usepackage{dcolumn}

\newcommand{\balpha}{{\mbox{\boldmath$\alpha$}}}
\newcommand{\bnabla}{{\mbox{\boldmath$\nabla$}}}

\newcommand{\be}{\begin{eqnarray}}
\newcommand{\ee}{\end{eqnarray}}
\newcommand{\la}{\langle}
\newcommand{\ra}{\rangle}

\newcommand{\bfx}{{\bf x}}
\newcommand{\bfy}{{\bf y}}
\newcommand{\bfk}{{\bf k}}
\newcommand{\bfr}{{\bf r}}
\newcommand{\bfp}{{\bf p}}
\newcommand{\bfq}{{\bf q}}
\newcommand{\rrp}{{\rm p}}

\newcommand{\rrk}{{\rm k}}
\newcommand{\veps}{\varepsilon}
\newcommand{\vare}{\varepsilon}
\newcommand{\eps}{\epsilon}
\newcommand{\beps}{{\mbox{\boldmath$\epsilon$}}}
\newcommand{\cross}[1]{#1\!\!\!/}

\begin{document}

\title{Radiative and correlation effects on
 the parity-nonconserving transition amplitude
in heavy alkaline atoms}

\author{V. M. Shabaev,$^{1,2}$ I. I. Tupitsyn,$^{1}$
K. Pachucki,$^{3}$
G. Plunien,$^{4}$  and V. A. Yerokhin$^{1,5}$}

\affiliation {$^{1}$Dept. of Physics, St.Petersburg State University,
Oulianovskaya 1, Petrodvorets, St.Petersburg 198504, Russia\\ 
$^2$  Max-Planck Institut f\"ur Physik Komplexer Systeme,
N\"othnitzer Stra{\ss}e 38, D-01187 Dresden, Germany\\
$^{3}$ Institute
of Theoretical Physics, Warsaw University, Ho\.za 69, 00-681, Warsaw, Poland\\
$^4$ Institut f\"ur Theoretische Physik, TU Dresden,
Mommsenstra{\ss}e 13, D-01062 Dresden, Germany \\
$^{5}$Center for Advanced Studies, St.Petersburg State Polytechnical University,
Politekhnicheskaya 29,
St.Petersburg 195251,
Russia}

\begin{abstract}

The complete gauge-invariant set of the one-loop QED corrections to the
parity-nonconserving (PNC)
amplitude in cesium  and  francium is evaluated to all orders
in $\alpha Z$ using a local form of the 
Dirac-Fock potential. 
The calculations are performed
in both length and velocity gauges for the absorbed photon and
the total binding QED correction is found
to be $-$0.27(3)\% for Cs and $-$0.28(5)\% for Fr.  
Moreover, a high-precision calculation of the electron-correlation 
and Breit-interaction effects
on the 7$s$-8$s$ PNC amplitude in francium
using a large-scale configuration-interaction Dirac-Fock method
is performed. 
The obtained results are employed to improve the
theoretical predictions
for the PNC transition amplitude in Cs and Fr.
Using an average value from two most accurate
measurements of the vector transition polarizability,
the weak charge of $^{133}$Cs is derived to  amount to
$ Q_W=-72.65(29)_{\rm exp}(36)_{\rm theor}$.
This value
deviates by $1.1\sigma$ from the prediction of the standard model.
The values of the $7s$-$8s$ PNC amplitude in $^{223}$Fr and $^{210}$Fr
are obtained to be $-$15.49(15) and $-$14.16(14), respectively,
in units of  i$\times 10^{-11}(-Q_W)/N$ a.u.

\end{abstract}
\maketitle

\section{Introduction}

Measurements of the parity nonconservation (PNC) effects  in atoms 
provide sensitive tests 
of the standard model (SM) 
 and impose constraints on physics beyond  it \cite{khr91,khr04}.
The 6$s$-7$s$ PNC amplitude in $^{133}$Cs 
\cite{bou74} 
remains one of the most effective tool for 
such investigations.
The measurement of this amplitude to a 0.3\% accuracy
\cite{wood97,ben99} has stimulated a reanalysis
of the theoretical predictions given
in Refs. \cite{dzu89,Johnson_90,blu00}. 
First, it was found 
\cite{der00,koz01,dzu01,der01,dzu02}
that the role of the Breit interaction had been
underestimated. 
Then, it was pointed out
 \cite{sush01} 
that the QED corrections may be comparable 
with the Breit corrections. The numerical evaluation
of the vacuum-polarization (VP) correction 
 \cite{joh01}
led to a 0.4\% increase
of the 6$s$-7$s$ PNC
amplitude in $^{133}$Cs, which resulted
in  a  2.2$\sigma$ deviation
of the weak charge of $^{133}$Cs
from the SM prediction.
This has triggered a great interest to calculations
of the complete one-loop QED corrections to the PNC amplitude.

While the VP contribution can easily be evaluated
to a high accuracy
within the Uehling approximation,
the calculation of the self-energy (SE) contribution is a much more demanding
problem (here and below we imply that the SE term embraces all one-loop
vertex diagrams as well). 
In the plane wave approximation, that corresponds
to zeroth order in $\alpha Z$,
it was derived
in Refs. \cite{mar83,lynn94}.
This correction, whose relative value equals 
to $-\alpha/(2\pi)$, is commonly 
 included in the definition of the nuclear weak charge. 
The $\alpha Z$-dependent part 
of the SE correction to the PNC matrix element between $s$ and $p$ states
was evaluated in Refs. \cite{kuch02,mil02}.
These calculations,
which are exact to first order in  $\alpha Z$ and partially include
higher-order binding effects, 
yield the binding SE  correction of
$-$0.9(1)\% \cite{kuch02,kuch03} and $-$0.85\% \cite{mil02}.
 The corresponding
total binding QED correction was found to amount to $-$0.5\% and $-$0.43\%,
 respectively. 
Despite this restored agreement with SM,
the  status of the QED correction to
PNC in $^{133}$Cs could not be considered
as resolved
until a complete $\alpha Z$-dependence calculation of
the SE correction to the 6$s$-7$s$ transition amplitude
 is accomplished. 
This is due to  the 
following reasons. First, in case of cesium $(Z=55)$ the parameter
$\alpha Z \approx 0.4$ is not small and, therefore, 
the higher-order corrections, which are beyond 
the  $A(\alpha Z) {\rm ln}(\lambda_{\rm C}/R_{\rm nuc})$ term  \cite{mil02},
 can be significant.
Second, because the calculations \cite{kuch02,mil02,kuch03}
are performed for the PNC matrix element only,
they do not include other SE diagrams 
which contribute to the 6$s$-7$s$ transition amplitude.
For instance, these calculations do not account for diagrams 
in which the virtual photon
embraces both the weak interaction
and the absorbed photon. 
Third, strictly speaking,
the PNC matrix element between the 
states of different  energies is not gauge invariant.
Despite the gauge-dependent part is suppressed by the small
energy difference \cite{mil02}, estimates of the uncertainty
in the definition of the PNC diagrams may fail due to
unphysical origin of the gauge-dependent terms.

The first step towards a complete $\alpha Z$-dependence 
calculation of the SE correction
was done in Ref. \cite{sap03}, where
the SE correction to the $2s$-$2p_{1/2}$ PNC matrix element
 in H-like ions
was evaluated. This matrix element
 was chosen to deal with 
the simplified gauge-invariant amplitude.
The results of that work agree with those of Refs.
\cite{kuch02,mil02,kuch03}. 
However, as was stressed there,
no claims can be made about the applicability of these
results to the 6$s$-7$s$ PNC transition in neutral cesium.

Finally, the whole gauge-invariant set of the one-loop QED corrections to 
the 6$s$-7$s$ PNC transition amplitude in cesium was evaluated
in Ref. \cite{sha05}. This calculation showed that the 
contributions of all SE diagrams
are of the same order of magnitude (in both length and velocity
gauges) and the final result arises through
a delicate cancellation of individual terms, none of which can be
neglected. The binding SE correction was obtained to amount to
$-$0.67(3)\% whereas the total binding QED correction is $-$0.27(3)\%.

Recently, the one-loop radiative corrections to the 6$s$-7$s$ PNC
amplitude in cesium were reevaluated by a semi-empirical method  \cite{fla05}.
In addition to the radiative correction to the weak matrix element, 
this method accounts for the
related corrections to the energy levels and to the electric dipole (E1)
 amplitude.
Despite it is intended to incorporate the radiative and correlation 
effects, it is unclear how the results obtained by this method are related
to those derived in the framework of the rigorous QED approach. 
The total binding QED correction obtained in  Ref. \cite{fla05}
amounts to $-$0.32(3)\%.

In the present paper we describe in detail 
the complete $\alpha Z$-dependence evaluation 
of the one-loop QED corrections to the PNC transition
amplitude in alkaline atoms and present the corresponding
numerical results for the 6$s$-7$s$ PNC amplitude
in cesium  \cite{sha05} and for the 7$s$-8$s$ PNC amplitude
in francium, which is going to be a subject of the PNC experiment,
as proposed in Ref. \cite{beh93}.
Moreover, we perform
a high-precision atomic structure calculation of the PNC transition
amplitude in francium using a large-scale configuration-interaction
Dirac-Fock (CI-DF) method and compare the results  with those from
Refs. \cite{dzu95,saf00}. 
The obtained contributions are combined with other terms
to improve the theoretical
predictions for the PNC transition amplitudes in Cs and Fr.

The relativistic units ($\hbar=c=1$) and the Heaviside charge
unit ($\alpha = e^2/(4\pi)$, $e<0$) are used throughout
the paper.

\section{QED corrections}

\subsection{Formulation}
A systematic derivation of the QED corrections in a fully relativistic
approach requires the use of perturbation theory starting with a one-electron
approximation in an effective local potential $V(r)$
   \begin{eqnarray}
 (-i\mbox{\boldmath $\alpha$}\cdot \mbox{\boldmath $\nabla$}
+\beta m+V(\bfx))\psi_{n}(\bfx)=
 \varepsilon_{n}\psi_{n}(\bfx)\,.
 \label{dirac}
    \end{eqnarray}
In neutral atoms, it is assumed that $V(r)$ includes the interaction
with the Coulomb field of the nucleus as well as
partly the electron-electron interaction.
The  interaction
of the electrons with the quantized electromagnetic field and the correlation
effects are  accounted for by the perturbation theory. 
In this way we obtain quantum electrodynamics in the Furry picture.

To derive formal expressions for the transition amplitude
we employ the method developed
in Ref. \cite{sha90} and described in detail in Ref. \cite{sha02}.
While this method is valid for arbitrary many-electron atom and
for arbitrary (single, degenerate, and quasidegenerate)
initial and final states, 
its formulation is especially simple for a one-electron atom (or an atom with one
electron over the closed shells) and for the case of single initial and
final states.

We consider the transition of the atom from the initial state
$a$ (which is 6$s$ for Cs and 7$s$ for Fr) 
to the final state $b$  (which is 7$s$ for Cs and 8$s$ for Fr)
accompanied by the absorption of a photon with momentum $\bfk$,
energy $k^0=|\bfk|$,
and polarization  $\eps^{\nu}=(0,\beps)$.
The transition amplitude is given by the formula \cite{sha90,sha02}
\be \label{transfin}
\tau &=&Z_3^{-1/2}\frac{1}{2\pi i}
\oint_{\Gamma_b}dE' \;\oint_{\Gamma_a}dE \; g_{b;\gamma,a}
(E',E)\nonumber\\
&&\times \Bigl[\frac{1}{2\pi i}\oint_{\Gamma_b}dE \; g_{bb}(E)\Bigr]^{-1/2}
\Bigl[\frac{1}{2\pi i}\oint_{\Gamma_a}dE \; g_{aa}(E)\Bigr]^{-1/2}
\,.
\ee
In the case under consideration (one electron over the closed shells),
the Green functions  $g_{b;\gamma,a}(E',E)$,
$g_{aa}(E)$, and $g_{bb}(E)$ are defined by
\be \label{g_b}
g_{b;\gamma,a}(E',E)\delta(E'-E-\omega)
&=&\int {d}{\bf x} {d}{\bf x'}
\psi^\dag_a({\bf x'})G_{\gamma}(E', {\bf x'};\omega;E, {\bf x})\gamma^0 
\psi_a({\bf x})\,,\\
g_{aa}(E)\delta(E'-E)
&=&\frac{2\pi }{i}\int {d}{\bf x} {d}{\bf x'}
\psi^\dag_a({\bf x'})G(E', {\bf x'};E, {\bf x})\gamma^0 \psi_a({\bf x})\,,
 \label{g_b1} \\
g_{bb}(E)\delta(E'-E)
&=&\frac{2\pi }{i}\int {d}{\bf x} {d}{\bf x'}
\psi^\dag_b({\bf x'})G(E', {\bf x'};E, {\bf x})\gamma^0 \psi_b({\bf x})\,.
 \label{g_b2}
\ee
Here
\be \label{gtrans23}
 G_{\gamma}(E',{\bfx'};\omega;E,\bfx)
&=&\frac{2\pi}{i}\frac{1}{(2\pi)^{3}}
\int_{-\infty}^{\infty}dx^{0}
dx^{\prime 0}\int d^4y
\exp{(iE'x^{\prime 0}
-iEx^{0}-i\omega y^0)}\nonumber\\
&&\times A^{\nu }(\bfy)
\la 0|T\psi(x^{\prime})j_{\nu}(y) 
\overline {\psi}(x)|0\ra\, 
\ee
is the Fourier transform of the Green function
describing the process,
$\psi(\bfx)$ is the electron-positron field operator in
the Heisenberg representation, $\overline{\psi} = \psi^{\dag}\gamma^0$,
$\gamma^0$ is the Dirac matrix,
\begin{eqnarray}\label{defA}
A^{\nu}(\bfx)=
 \frac{{\eps}^{\nu} \exp{(i{\bf k}\cdot{\bf x})}}
{\sqrt{2k^{0}(2\pi )^3}}\,
\end{eqnarray}
is the wave function of the absorbed photon, and 
\be \label{gtrans24}
 G(E',\bfx';E,\bfx)=\frac{1}{(2\pi)^{2}}
\int_{-\infty}^{\infty}dx^{0}
dx^{\prime 0}
\exp{(iE'x^{\prime 0}
-iE x^{0})} \la 0|T\psi(x^{\prime}) 
\overline {\psi}(x)|0\ra\, 
\ee
is the Fourier transform of the Green function
describing the atom.
The contours $\Gamma_a$ and $\Gamma_b$ surround the
poles corresponding to the initial and final levels and
keep outside all other singularities of the Green functions.
It is assumed that they are oriented anticlockwise.
$Z_3$ is the renormalization constant for the photon wave
function and the factors
$\Bigl[\frac{1}{2\pi i}\oint_{\Gamma_a}dE \; g_{aa}(E)\Bigr]^{-1/2}$
and 
$\Bigl[\frac{1}{2\pi i}\oint_{\Gamma_b}dE \; g_{bb}(E)\Bigr]^{-1/2}$
serve as the normalization factors for the electron wave functions
of the states $a$ and $b$, respectively.
The  Green functions $G$ and $G_{\gamma}$ are
constructed by perturbation theory after the transition to
the interaction representation and using Wick's theorem.
The Feynman rules for $G$ and  $G_{\gamma}$ are given
in Ref. \cite{sha02}.

To the lowest order, the PNC transition amplitude is described
by diagrams presented in Fig~\ref{fig:pnc}. Denoting the contribution
to $g_{b;\gamma,a}(E',E)$ from these diagrams 
by $g^{(0)}_{b;\gamma,a}(E',E)$ and taking into account
that
$g^{(0)}_{aa}=1/(E-\veps_a)$ and $g^{(0)}_{bb}=1/(E-\veps_b)$
and, therefore,
 the normalization factors in formula 
(\ref{transfin}) are equal to 1,
we obtain
\be \label{transfin0}
\tau^{(0)}&=&\frac{1}{2\pi i}
\oint_{\Gamma_b}dE' \;\oint_{\Gamma_a}dE \; g^{(0)}_{b;\gamma,a}
(E',E)\,.
\ee
According to the Feynman rules \cite{sha02} and definition 
(\ref{g_b}),
 we have
\be \label{g_b_0}
g^{(0)}_{b;\gamma,a}(E',E)&=&\frac{i}{2\pi}\frac{1}{E'-\veps_b}
\sum_{n}\frac{\la b |e\alpha^{\nu}A_{\nu}|n\ra \la n |H_W|a\ra}
{E-\veps_n}\frac{1}{E-\veps_a}\nonumber\\
& &+ \frac{i}{2\pi}\frac{1}{E'-\veps_b}
\sum_{n}\frac{\la b |H_w|n\ra \la n |e\alpha^{\nu}A_{\nu}
|a\ra}
{E'-\veps_n}\frac{1}{E-\veps_a}\,.
\ee
Here 
\be \label{h_w}
H_W=-(G_F/\sqrt{8})Q_W \rho_{N}(r)\gamma_5 
\ee
is the nuclear spin-independent weak-interaction Hamiltonian  \cite{khr91},
$G_F$ is the Fermi constant, $\gamma_5$ and
$\alpha^{\nu}\equiv \gamma^0\gamma^{\nu}=(1,\balpha)$
 are the Dirac matrices, and
$\rho_{N}$  is the nuclear
weak-charge density normalized to unity.
Substituting expression (\ref{g_b_0}) into equation (\ref{transfin0})
and taking into account that, for a non-Coulomb potential $V(r)$, 
there is no states
of different parity but of the same energy, we obtain 
 \be \label{t_b_0}
\tau^{(0)} =-
\sum_{n}\Bigl[
\frac{\la b |e\alpha^{\nu}A_{\nu}|n\ra \la n |H_W|a\ra}
{\veps_a-\veps_n}+ \frac{\la b |H_w|n\ra \la n |e\alpha^{\nu}A_{\nu}
|a\ra}
{\veps_b-\veps_n}\Bigr]\,.
\ee
We note here that the case of degenerate levels, which takes place for
the pure Coulomb potential, can be considered employing the related
formulas from Ref. \cite{sha02}.

The one-loop SE corrections to the PNC transition amplitude
are defined by diagrams presented in Fig.~\ref{fig:sepnc}.
To derive the formal expressions  for these corrections, 
one has to expand formula (\ref{transfin}) to the next-to-leading 
order:
\begin{eqnarray}\label{first1}
\tau^{(1)}&=&
\frac{1}{2\pi i}\;\Bigl\{
\oint_{\Gamma_b}dE' \; \oint_{\Gamma_a}dE
 \; g_{b;\gamma,a}^{(1)}(E',E)\nonumber\\
&&-\frac{1}{2}
\oint_{\Gamma_b}dE' \; \oint_{\Gamma_a}dE
 \; g_{b;\gamma,a}^{(0)}(E',E)\Bigl[
\frac{1}{2\pi i}\oint_{\Gamma_b} dE\; g_{bb}^{(1)}(E)
+\frac{1}{2\pi i}\oint_{\Gamma_a} dE\; g_{aa}^{(1)}(E)
\Bigr] \Bigr\}\,.
\end{eqnarray}
Let us consider the derivation of the contributions from
the diagrams ``a'' and ``c''. According to the Feynman rules \cite{sha02},
we have 
\be \label{g_a_1}
g^{(1,a)}_{b;\gamma,a}(E',E)&=&\frac{i}{2\pi}\frac{1}{E'-\veps_b}
\sum_{n_1 n_2}\frac{\la b |\Sigma(E')|n_1\ra \la n_1 |
e\alpha^{\nu}A_{\nu}|n_2\ra \la n_2 |H_W|a\ra}
{(E'-\veps_{n_1})(E-\veps_{n_2})}
\frac{1}{E-\veps_a}\,,\\
g^{(1,c)}_{b;\gamma,a}(E',E)&=&\frac{i}{2\pi}\frac{1}{E'-\veps_b}
\sum_{n_1 n_2}\frac{\la b |\Sigma(E')|n_1\ra \la n_1 |
H_W|n_2\ra \la n_2 | e\alpha^{\nu}A_{\nu}
|a\ra}
{(E'-\veps_{n_1})(E'-\veps_{n_2})}
\frac{1}{E-\veps_a}\,.
 \label{g_c_1}
\ee
Here the SE operator is defined as
\begin{eqnarray}
\la c|\Sigma(E)|d\ra  \equiv  \frac{i}{2\pi} \int_{-\infty}^{\infty}
d\omega \;
\sum_{n}\frac{\la cn|
I(\omega)|n d\ra}{E-\omega-u\veps_n} \,,
\end{eqnarray}
where $I(\omega) \equiv  e^2\alpha^{\mu}\alpha^{\nu}D_{\mu \nu}(\omega)$, 
$D_{\mu \nu}(\omega)$ is the photon propagator defined as
in Ref. \cite{sha02},
and $u=1-i0$ ensures the correct position of poles of the electron
propagators with respect to the integration contour.
Taking into account that, for a non-Coulomb potential, 
the energy $\veps_{n_2}$ in Eq. (\ref{g_a_1})
is never equal to $\veps_a$ and 
the energy $\veps_{n_2}$ in Eq. (\ref{g_c_1})
is never equal to $\veps_b$, 
we obtain
\be \label{gg_a}
\oint_{\Gamma_b}dE' \;\oint_{\Gamma_a}dE \; g^{(1,a)}_{b;\gamma,a}
(E',E)&=&-2\pi i
\Bigl[\sum_{n_1,n_2}^{n_1\ne b}
\frac{\la b |\Sigma(\veps_b)|n_1\ra \la n_1 |
e\alpha^{\nu}A_{\nu}|n_2\ra \la n_2 |H_W|a\ra}
{(\veps_b-\veps_{n_1})
(\veps_a-\veps_{n_2})}\nonumber\\
&& + 
\sum_{n}
\frac{\la b |\Sigma^{\prime}
(\veps_b)|b \ra \la b|
e\alpha^{\nu}A_{\nu}|n\ra \la n |H_W|a\ra}
{\veps_a-\veps_{n}}
\Bigr]\,,\\
\oint_{\Gamma_b}dE' \;\oint_{\Gamma_a}dE \; g^{(1,c)}_{b;\gamma,a}
(E',E)&=&-2\pi i
\Bigl[\sum_{n_1,n_2}^{n_1\ne b}
\frac{\la b |\Sigma(\veps_b)|n_1\ra \la n_1 |H_W
|n_2\ra \la n_2 | e\alpha^{\nu}A_{\nu}|a\ra}
{(\veps_b-\veps_{n_1})
(\veps_b-\veps_{n_2})}\nonumber\\
&& + 
\sum_{n}
\frac{\la b |\Sigma^{\prime}
(\veps_b)|b \ra \la b|H_W|n\ra \la n |e\alpha^{\nu}A_{\nu}|a\ra}
{\veps_b-\veps_{n}}\nonumber\\
&& -
\sum_{n}
\frac{\la b |\Sigma
(\veps_b)|b \ra \la b|H_W|n\ra \la n |e\alpha^{\nu}A_{\nu}|a\ra}
{(\veps_b-\veps_{n})^2}\Bigr]\,, \label{gg_c}
\ee
where  $\Sigma'(E)=d\Sigma(E)/dE$. The contributions containing
$ \la b |\Sigma^{\prime} (\veps_b)|b \ra$ should be considered together
with the second term in equation (\ref{first1}).
Taking into account that 
\be 
\frac{1}{2\pi i}\oint_{\Gamma_b} dE \; g_{bb}^{(1)}(E) 
=\frac{1}{2\pi i}\oint_{\Gamma_b} dE \;
\frac{ \la b|\Sigma(E)|b\ra}{(E-\veps_b)^2}
=\la b|\Sigma'(\veps_b)|b\ra\,,
\ee
we obtain
\be
&&-\frac{1}{2}
\Bigl[ \oint_{\Gamma_b}dE' \;\oint_{\Gamma_a}dE \; g^{(0)}_{b;\gamma,a}
(E',E)\Bigr]\Bigl[\frac{1}{2\pi i}\oint_{\Gamma_b} dE \; 
\Delta g_{bb}^{(1)}(E)\Bigr]\nonumber\\
&&\;\;\;\;\;\;\;\;\;\;\;=\frac{1}{2}2\pi i 
\sum_{n}\Bigl[
\frac{\la b |e\alpha^{\nu}A_{\nu}|n\ra \la n |H_W|a\ra}
{\veps_a-\veps_n}
+ \frac{\la b |H_w|n\ra \la n |e\alpha^{\nu}A_{\nu}
|a\ra}
{\veps_b-\veps_n}\Bigr]
 \la b|\Sigma^{\prime}(\veps_b)|b\ra\,.
\ee
Adding this contribution to the terms (\ref{gg_a}) and (\ref{gg_c}),
we obtain 
\be \label{tt_a}
\tau^{(1,a)}&=&-
\Bigl[\sum_{n_1,n_2}^{n_1\ne b}
\frac{\la b |\Sigma(\veps_b)|n_1\ra \la n_1 |
e\alpha^{\nu}A_{\nu}|n_2\ra \la n_2 |H_W|a\ra}
{(\veps_b-\veps_{n_1})
(\veps_a-\veps_{n_2})}\nonumber\\
&& + \frac{1}{2}
\sum_{n}
\frac{\la b |\Sigma^{\prime}
(\veps_b)|b \ra \la b|
e\alpha^{\nu}A_{\nu}|n\ra \la n |H_W|a\ra}
{\veps_a-\veps_{n}}
\Bigr]\,,\\
\tau^{(1,c)}&=&-
\Bigl[\sum_{n_1,n_2}^{n_1\ne b}
\frac{\la b |\Sigma(\veps_b)|n_1\ra \la n_1 |H_W
|n_2\ra \la n_2 | e\alpha^{\nu}A_{\nu}|a\ra}
{(\veps_b-\veps_{n_1})
(\veps_b-\veps_{n_2})}\nonumber\\
&& + \frac{1}{2}
\sum_{n}
\frac{\la b |\Sigma^{\prime}
(\veps_b)|b \ra \la b|H_W|n\ra \la n |e\alpha^{\nu}A_{\nu}|a\ra}
{\veps_b-\veps_{n}}\nonumber\\
&& -
\sum_{n}
\frac{\la b |\Sigma
(\veps_b)|b \ra \la b|H_W|n\ra \la n |e\alpha^{\nu}A_{\nu}|a\ra}
{(\veps_b-\veps_{n})^2}\Bigr]\,. \label{tt_c}
\ee
Similar calculations yield
\be \label{tt_b}
\tau^{(1,b)}&=&-
\Bigl[\sum_{n_1,n_2}^{n_2\ne a}
\frac{\la b |H_W|n_1\ra \la n_1 |
e\alpha^{\nu}A_{\nu}|n_2\ra \la n_2 |
\Sigma(\veps_a)|a\ra}
{(\veps_b-\veps_{n_1})
(\veps_a-\veps_{n_2})}\nonumber\\
&& + \frac{1}{2}
\sum_{n}
\frac{\la b |H_W|n \ra \la n|
e\alpha^{\nu}A_{\nu}|a\ra \la a |
\Sigma^{\prime}
(\veps_a)|a\ra}
{\veps_b-\veps_{n}}
\Bigr]\,,
\ee
\be
\tau^{(1,d)}&=&-
\Bigl[\sum_{n_1,n_2}^{n_1\ne b}
\frac{\la b | e\alpha^{\nu}A_{\nu}
|n_1\ra \la n_1 |H_W
|n_2\ra \la n_2 |
\Sigma(\veps_a)
|a\ra}
{(\veps_a-\veps_{n_1})
(\veps_a-\veps_{n_2})}\nonumber\\
&& + \frac{1}{2}
\sum_{n}
\frac{\la b |e\alpha^{\nu}A_{\nu}
|n \ra \la n|H_W|a\ra \la a | \Sigma^{\prime}(\veps_a)
|a\ra}
{\veps_a-\veps_{n}}\nonumber\\
&& -
\sum_{n}
\frac{\la b | e\alpha^{\nu}A_{\nu}
|n \ra \la n|H_W|a\ra \la a | \Sigma (\veps_a)
|a\ra}
{(\veps_a-\veps_{n})^2}\Bigr]\, \label{tt_d}\,,
\ee
\be
\tau^{(1,e)}=-\sum_{n_1,n_2}^{n_1\ne b}
\frac{\la b | e\alpha^{\nu}A_{\nu}
|n_1\ra \la n_1 |\Sigma(\veps_a)
|n_2\ra \la n_2 |H_W
|a\ra}
{(\veps_a-\veps_{n_1})
(\veps_a-\veps_{n_2})}\,,
\ee
\be
\tau^{(1,f)}=-\sum_{n_1,n_2}^{n_2\ne a}
\frac{\la b |H_W|n_1\ra \la n_1 |\Sigma(\veps_b)|n_2\ra \la n_2 |
e\alpha^{\nu}A_{\nu}
|a\ra}
{(\veps_b-\veps_{n_1})
(\veps_b-\veps_{n_2})}\,,
\ee
\be
\label{tt_g}
\tau^{(1,g)}&=&-\frac{i}{2\pi} \int_{-\infty}^{\infty} d\omega
\sum_{n,n_1,n_2}
\frac{\la n_1|
e\alpha^{\nu}A_{\nu}
|n_2\ra
\la n | H_W|a\ra}{(\vare_a-\vare_{n})}\nonumber\\
&&\times
\frac{\la b n_2|I(\omega)|n_1 n\ra}
{[\vare_b-\omega-u\vare_{n_1}]
 [\vare_a-\omega-u\vare_{n_2}]}\,,
\ee
\be
\label{tt_h}
\tau^{(1,h)}
&=&-\frac{i}{2\pi} \int_{-\infty}^{\infty} d\omega
\sum_{n,n_1,n_2}
\frac{\la b | H_W|n\ra 
\la n_1|
e\alpha^{\nu}A_{\nu}
|n_2\ra}{(\vare_b-\vare_{n})}\nonumber\\
&&\times \frac{\la n n_2|I(\omega)|n_1 a\ra}
{ [\vare_b-\omega-u\vare_{n_1}]
 [\vare_a-\omega-u\vare_{n_2}]}\,,
\ee
\be
\label{tt_i}
\tau^{(1,i)}
&=&-\frac{i}{2\pi} \int_{-\infty}^{\infty} d\omega
\sum_{n,n_1,n_2}
\frac{\la n_1|H_W|n_2\ra
\la n | e\alpha^{\nu}A_{\nu}
|a\ra}{(\vare_b-\vare_{n})}\nonumber\\
&&\times\frac{\la b n_2|I(\omega)|n_1 n\ra}
{[\vare_b-\omega-u\vare_{n_1}]
 [\vare_b-\omega-u\vare_{n_2}]}\,,
\ee
\be
\label{tt_j}
\tau^{(1,j)}&=&-\frac{i}{2\pi} \int_{-\infty}^{\infty} d\omega
\sum_{n,n_1,n_2}
\frac{\la b | e\alpha^{\nu}A_{\nu}|n\ra 
\la n_1|H_W|n_2\ra}{(\vare_a-\vare_{n})}
\nonumber\\
&&\times \frac{\la n n_2|I(\omega)|n_1 a\ra}
{ [\vare_a-\omega-u\vare_{n_1}]
[\vare_a-\omega-u\vare_{n_2}]}\,,
\ee
\be
\label{tt_k}
\tau^{(1,k)}
&=&-\frac{i}{2\pi} \int_{-\infty}^{\infty} d\omega
\sum_{n_1,n_2,n_3}
\frac{\la b n_2|I(\omega)|n_1 a\ra}{[\vare_b-\omega-u\vare_{n_1}]}
\nonumber\\
&&\times\frac{\la n_1|
 e\alpha^{\nu}A_{\nu}|n_3\ra
\la n_3 | H_W|n_2\ra}
{[\vare_a-\omega-u\vare_{n_3}]
 [\vare_a-\omega-u\vare_{n_2}]}\,,
\ee
\be
\label{tt_l}
\tau^{(1,l)}
&=&-\frac{i}{2\pi} \int_{-\infty}^{\infty} d\omega
\sum_{n_1,n_2,n_3}
\frac{\la b n_2|I(\omega)|n_1 a\ra}{[\vare_b-\omega-u\vare_{n_1}]}\nonumber\\
&&\times\frac{ \la n_1|H_W|n_3\ra
\la n_3 |  e\alpha^{\nu}A_{\nu}
|n_2\ra}
{[\vare_b-\omega-u\vare_{n_3}]
 [\vare_a-\omega-u\vare_{n_2}]}\,.
\ee
Taking into account the corresponding diagrams with the mass counterterm 
results in the replacement $\Sigma(E)\rightarrow \Sigma_{\rm R}(E)=
\Sigma(E)-\gamma^0\delta m$.

Since the wave length of the absorbed photon is much larger than
the atomic size, one can use the dipole approximation.
It means the replacement $\exp{(i{\bf k}\cdot {\bf
    x})}\rightarrow 1$ in the photon wave function and, therefore,
$ e\alpha^{\nu}A_{\nu} \rightarrow |e|(\balpha
\cdot\beps)/\sqrt{2k^0(2\pi)^3}$ in formulas (\ref{t_b_0}),
(\ref{tt_a})-(\ref{tt_l}). 
Within this approximation, 
the corresponding formulas in the length gauge are obtained
by replacing $\balpha$ with $\bfr$ 
in all vertices
corresponding to the photon absorption and by multiplying
the  amplitude (\ref{t_b_0})
with the factor 
$i(E_b-E_a)$, where 
 $E_a=\veps_a+\la a|\Sigma_{\rm R}(\veps_a)|a\ra$ and 
$E_b=\veps_b+\la b|\Sigma_{\rm R}(\veps_b)|b\ra$,
and  the amplitudes  (\ref{tt_a})-(\ref{tt_l})
with the factor $i(\veps_b-\veps_a)$.
This prescription can
 be derived  from equations
(\ref{t_b_0}), (\ref{tt_a})
-(\ref{tt_l}) employing
 the commutation relation $\balpha =i[h_{\rm D},\bfr]$, where $h_{\rm D} =
-i\mbox{\boldmath $\alpha$}\cdot \mbox{\boldmath $\nabla$}
+\beta m+V(r)$ is the Dirac Hamiltonian. Alternatively,
one can get it
using Eq. (205) of Ref. \cite{sha02}
and the equal-time 
commutation relations for the field operators in the Heisenberg
representation.

The theoretical and experimental results for the PNC amplitude 
in alkaline atoms
are generally presented in terms of the $E_{\rm PNC}$ amplitude
which is defined as the matrix element of the z component  
of the atomic electric-dipole moment 
 between the initial ($a$)
and final ($b$) $s$ states with the angular momentum projections
$m_a=m_b=1/2$.
It is related to the $\tau$ amplitude by the equation
\be \label{def_E_PNC}
E_{\rm PNC}=i\tau[e\alpha^{\nu}A_{\nu} \rightarrow e\alpha_z]/(E_b-E_a)
 = \tau[e\alpha^{\nu}A_{\nu} \rightarrow -d_z]\,,
\ee 
where the $s$ states $a$ and $b$ have the angular momentum projections
$m_a=m_b=1/2$, 
$E_a$ and $E_b$ are their total energies, and 
$d_z=ez$ is the $z$ component of the dipole moment operator ($e<0$).
To zeroth order, one easily finds
\begin{eqnarray}\label{zero}
E_{\rm PNC} = \sum_n\Bigl[\frac{\la b|d_z|n\ra \la n|H_W|a\ra}{\vare_a-\vare_n}
+\frac{\la b|H_W|n\ra \la n|d_z|a\ra}{\vare_b-\vare_n}\Bigr]\,.
\end{eqnarray}
The one-loop SE correction 
 is given by the sum of the following terms:
\begin{eqnarray}\label{se_a}
\delta E_{\rm PNC}^{\rm a}&=& \sum_{n_1,n_2}^{(n_1\ne b)}
\frac{\la b|\Sigma_{\rm R}(\vare_b)|n_1\ra \la n_1|d_z|n_2\ra \la n_2 | H_W|a\ra}
{(\vare_b-\vare_{n_1}) (\vare_a-\vare_{n_2})}\nonumber\\
&&+\frac{1}{2}\sum_{n}
\frac{\la b|\Sigma'_{\rm R}(\vare_b)|b\ra \la b|d_z|n\ra \la n | H_W|a\ra}
{ (\vare_a-\vare_{n})}\,,
\end{eqnarray}
\begin{eqnarray}\label{se_b}
\delta E_{\rm PNC}^{\rm b}&=& \sum_{n_1,n_2}^{(n_2\ne a)}
\frac{\la b|H_W|n_1\ra \la n_1|d_z|n_2\ra \la n_2 |\Sigma_{\rm R}(\vare_a)|a\ra}
{(\vare_b-\vare_{n_1}) (\vare_a-\vare_{n_2})}\nonumber\\
&&+\frac{1}{2}\sum_{n}
\frac{\la b|H_W|n\ra \la n|d_z|a\ra \la a | \Sigma'_{\rm R}(\vare_a)
|a\ra}{ (\vare_b-\vare_{n})}\,,
\end{eqnarray}
\begin{eqnarray}\label{se_c}
\delta E_{\rm PNC}^{\rm c}&=& \sum_{n_1,n_2}^{(n_1\ne b)}
\frac{\la b|\Sigma_{\rm R}(\vare_b)|n_1\ra \la n_1|H_W|n_2\ra \la n_2 | d_z|a\ra}
{(\vare_b-\vare_{n_1}) (\vare_b-\vare_{n_2})}\nonumber\\
&&+\frac{1}{2}\sum_{n}
\frac{\la b|\Sigma'_{\rm R}(\vare_b)|b\ra \la b|H_W|n\ra \la n | d_z|a\ra}
{ (\vare_b-\vare_{n})}\nonumber\\
&&-\sum_{n}
\frac{\la b|\Sigma_{\rm R}(\vare_b)|b\ra \la b|H_W|n\ra \la n | d_z|a\ra}
{ (\vare_b-\vare_{n})^2}\,,
\end{eqnarray}
\begin{eqnarray}\label{se_d}
\delta E_{\rm PNC}^{\rm d}&=& \sum_{n_1,n_2}^{(n_2\ne a)}
\frac{\la b|d_z|n_1\ra \la n_1|H_W|n_2\ra \la n_2 |\Sigma_{\rm R}(\vare_a)|a\ra}
{(\vare_a-\vare_{n_1}) (\vare_a-\vare_{n_2})}\nonumber\\
&&+\frac{1}{2}\sum_{n}
\frac{\la b|d_z|n\ra \la n|H_W|a\ra \la a | \Sigma'_{\rm R}(\vare_a)
|a\ra}{ (\vare_a-\vare_{n})}\nonumber\\
&&-\sum_{n}
\frac{\la b|d_z|n\ra \la n|H_W|a\ra \la a | \Sigma_{\rm R}(\vare_a)
|a\ra}{ (\vare_a-\vare_{n})^2}\,,
\end{eqnarray}
\begin{eqnarray} \label{se_e}
\delta E_{\rm PNC}^{\rm e}= \sum_{n_1,n_2}
\frac{\la b|d_z|n_1\ra \la n_1| \Sigma_{\rm R}(\vare_a)|n_2\ra \la n_2 | H_W|a\ra}
{(\vare_a-\vare_{n_1}) (\vare_a-\vare_{n_2})}\,,
\end{eqnarray}
\begin{eqnarray}\label{se_f}
\delta E_{\rm PNC}^{\rm f}= \sum_{n_1,n_2}
\frac{\la b|H_W|n_1\ra \la n_1| \Sigma_{\rm R}(\vare_b)|n_2\ra \la n_2 | d_z|a\ra}
{(\vare_b-\vare_{n_1}) (\vare_b-\vare_{n_2})}\,,
\end{eqnarray}
\begin{eqnarray}\label{se_g}
\delta E_{\rm PNC}^{\rm g}&=&\frac{i}{2\pi} \int_{-\infty}^{\infty} d\omega
\sum_{n,n_1,n_2}
\frac{\la n_1|d_z|n_2\ra
\la n | H_W|a\ra}{(\vare_a-\vare_{n})}\nonumber\\
&&\times
\frac{\la b n_2|I(\omega)|n_1 n\ra}
{[\vare_b-\omega-u\vare_{n_1}]
 [\vare_a-\omega-u\vare_{n_2}]}\,,
\end{eqnarray}
\begin{eqnarray}\label{se_h}
\delta E_{\rm PNC}^{\rm h}&=&\frac{i}{2\pi} \int_{-\infty}^{\infty} d\omega
\sum_{n,n_1,n_2}
\frac{\la b | H_W|n\ra 
\la n_1|d_z|n_2\ra}{(\vare_b-\vare_{n})}\nonumber\\
&&\times \frac{\la n n_2|I(\omega)|n_1 a\ra}
{ [\vare_b-\omega-u\vare_{n_1}]
 [\vare_a-\omega-u\vare_{n_2}]}\,,
\end{eqnarray}
\begin{eqnarray}\label{se_i}
\delta E_{\rm PNC}^{\rm i}&=&\frac{i}{2\pi} \int_{-\infty}^{\infty} d\omega
\sum_{n,n_1,n_2}
\frac{\la n_1|H_W|n_2\ra
\la n | d_z|a\ra}{(\vare_b-\vare_{n})}\nonumber\\
&&\times\frac{\la b n_2|I(\omega)|n_1 n\ra}
{[\vare_b-\omega-u\vare_{n_1}]
 [\vare_b-\omega-u\vare_{n_2}]}\,,
\end{eqnarray}
\begin{eqnarray}\label{se_j}
\delta E_{\rm PNC}^{\rm j}&=&\frac{i}{2\pi} \int_{-\infty}^{\infty} d\omega
\sum_{n,n_1,n_2}
\frac{\la b | d_z|n\ra 
\la n_1|H_W|n_2\ra}{(\vare_a-\vare_{n})}
\nonumber\\
&&\times \frac{\la n n_2|I(\omega)|n_1 a\ra}
{ [\vare_a-\omega-u\vare_{n_1}]
[\vare_a-\omega-u\vare_{n_2}]}\,,
\end{eqnarray}
\begin{eqnarray}\label{se_k}
\delta E_{\rm PNC}^{\rm k}&=&\frac{i}{2\pi} \int_{-\infty}^{\infty} d\omega
\sum_{n_1,n_2,n_3}
\frac{\la b n_2|I(\omega)|n_1 a\ra}{[\vare_b-\omega-u\vare_{n_1}]}
\nonumber\\
&&\times\frac{\la n_1|d_z|n_3\ra
\la n_3 | H_W|n_2\ra}
{[\vare_a-\omega-u\vare_{n_3}]
 [\vare_a-\omega-u\vare_{n_2}]}\,,
\end{eqnarray}
\begin{eqnarray}\label{se_l}
\delta E_{\rm PNC}^{\rm l}&=&\frac{i}{2\pi} \int_{-\infty}^{\infty} d\omega
\sum_{n_1,n_2,n_3}
\frac{\la b n_2|I(\omega)|n_1 a\ra}{[\vare_b-\omega-u\vare_{n_1}]}\nonumber\\
&&\times\frac{ \la n_1|H_W|n_3\ra
\la n_3 | d_z|n_2\ra}
{[\vare_b-\omega-u\vare_{n_3}]
 [\vare_a-\omega-u\vare_{n_2}]}\,.
\end{eqnarray}

According to Eq. (\ref{def_E_PNC}),
the corresponding expressions in the velocity gauge are 
obtained by the replacement  $d_z \rightarrow -ie\alpha_z/(E_b-E_a)$,
where the energies $E_a$ and $E_b$ include the SE corrections.
In addition to the replacement
$d_z \rightarrow -ie\alpha_z/(\vare_b-\vare_a)$
in Eqs. (\ref{zero})-(\ref{se_l}), it yields the contribution 
\begin{eqnarray} \label{se_add}
\delta E_{\rm PNC}^{\rm add}=
-\frac{\la b|\Sigma_{\rm R}(\vare_b)|b\ra - \la a|\Sigma_{\rm R}(\vare_a)|a\ra}
{\vare_b-\vare_a}E_{\rm PNC}\,,
\end{eqnarray}
which results from the expansion
\be 
\frac{1}{E_b-E_a} 
\approx \frac{1}{\veps_b-\veps_a}\Bigl[1
-\frac{\la b|\Sigma_{\rm R}(\vare_b)|b\ra - \la a|\Sigma_{\rm
    R}(\vare_a)|a\ra}
{(\veps_b-\veps_a)}\Bigr]\,.
\ee
It can be shown that the sum of contributions
(\ref{se_i})-(\ref{se_l}) is the same in the
length and the velocity gauge. 
Because of the gauge invariance of the total SE
correction,
the same is valid for
the sum of the other terms, Eqs. 
(\ref{se_a})-(\ref{se_h}) and (\ref{se_add}).

Formulas (\ref{se_a})-(\ref{se_add})
 contain ultraviolet and infrared divergences.
To cancel the ultraviolet divergences, we 
expand 
 contributions (\ref{se_a})-(\ref{se_f}) into
zero-, one-, and many-potential terms
and  contributions (\ref{se_g})-(\ref{se_j})
into zero- and many-potential terms.
The ultraviolet divergencies are present only in the
zero- and one-potential terms. They are removed
analytically by calculating these terms in the momentum space
 (for details, we refer to Refs. \cite{sny91,yer99,sap02}). 
For the standard zero- and
one-potential terms we employ the equations given in Ref.
\cite{yer99} whereas  the corresponding  expression for the zero-potential 
PNC term is presented in the Appendix. 
The many-potential terms are evaluated in configuration space
employing the Wick rotation in the complex $\omega$ plane. 
The infrared divergences, which occur in contributions 
(\ref{se_a})-(\ref{se_d}) and (\ref{se_k})-(\ref{se_l}),
are regularized by introducing a nonzero photon mass and cancelled
analytically.

The  expressions for the VP corrections, which
do not contain any insertions with the external photon line or the
weak interaction attached to the electron loop,  
are obtained from Eqs. (\ref{se_a})-(\ref{se_f}) by the replacement 
of the SE operator with the VP potential. The 
other VP corrections
will not be considered  here, since their contribution
is negligible. 
To a high accuracy, the VP potential is
determined by the Uehling term, which corresponds to first 
nonzero term in the expansion of the vacuum loop in powers of the
Coulomb potential.
 The renormalized expression for 
the Uehling potential is 
 \begin {eqnarray} \label{uehlexpr}
U_{\rm Uehl}(r)&=&-\alpha Z
\frac{2\alpha}{3\pi}\int\limits_0^\infty dr'\; 4\pi r'\rho
(r')
\int\limits_1^\infty dt \;
(1 +\frac{1}{2t^2})
\frac{\sqrt{t^2-1}}{t^{2}}\nonumber \\
&&\times \frac{[\exp{(-2m|r-r'|t)}-\exp{(-2m(r+r')t)}]}
{4mrt} \,,
\end{eqnarray}
where $\rho(r)$ is the nuclear charge density, normalized to unity.
To account for the screening effect on the Uehling potential,
one should replace $Z\rho(r)$ by  $Z\rho(r) -(Z-1) \rho_{\rm core}(r)$,
where $\rho_{\rm core}(r)$ is the charge density of the core electrons,
normalized to unity.
The higher-order one-loop VP potential, so-called
Wichmann-Kroll term, can be evaluated for the point-charge nucleus
using approximate formulas derived in Ref. \cite{fain91}.

\subsection{Local Dirac-Fock potential}

Since the energy intervals between the levels
6$s$, 6$p_{1/2}$, 7$s$, and 7$p_{1/2}$  in Cs
and the levels 7$s$, 7$p_{1/2}$, 8$s$, and 8$p_{1/2}$ in Fr
are very small,
to get reliable results for the
transition amplitudes under consideration,
one needs to use a local potential $V(r)$ that reproduces
 energies and wave functions of these states on the 
Dirac-Fock (DF) accuracy level or better.
We construct such a potential by inverting the radial Dirac equation
with the radial wave function obtained by solving the 
DHF equation
with the code of Ref. \cite{bra77}.

The radial DF equations  have the form \cite{bra77}
\begin{equation}
\left \{
\begin{array}{llll}
& \displaystyle -\, \left ( \frac{d}{dr} \,-\,
\frac{\kappa_a}{r} \right ) F_a \,+\,
\Bigl(V_{\rm C}+\frac{Y_a(r)}{r}\Bigr) \, G_a +m\, G_a &=& 
\displaystyle \varepsilon_a \, G_a \,-\,
\frac{X_a^F}{r}
\\ \rule{0pt}{7mm} \displaystyle
& \displaystyle  \, \left ( \frac{d}{dr} \,+\, \frac{\kappa_a}{r} \right )
G_a \,+\, \Bigl(V_{\rm C}+\frac{Y_a(r)}{r}\Bigr) \, F_a  \,-\,m \,F_a &=& \displaystyle
\varepsilon_a \,F_a \,-\, \frac{X_a^G}{r} \,.
\end{array} \right .
\label{hfd1}
\end{equation}
Here $G_a/r=g_a$ and $F_a/r=f_a$ are the large and small
radial components of the Dirac wave function of the $a$ shell electron,
 $\veps_a$ is the one-electron
 energy, $\kappa_a=(-1)^{j+l+1/2}(j+1/2)$ is the relativistic 
quantum number, $V_{\rm C}$ is the Coulomb potential induced by the nucleus,
and $Y_a(r)/r$ is the screening potential.
The functions $X_a^G$ and $X_a^F$ consist of two parts.
The first part is the result of the action of the exchange-interaction
 operator on the radial wave functions $G_a$ and $F_a$. 
The second part is the contribution from the non-diagonal Lagrangian
multipliers, which provide the orthogonality of the radial wave
functions corresponding to different values of the principal
quantum number $n_a$ but the same $\kappa_a$.
The functions  $X_a^G$ and $X_a^F$ are calculated self-consistently 
from the DF equations employng the radial wave functions obtained
at the previous iteration step.

Let us consider the Dirac equation for the $a$ shell electron  with a local
potential $V_a(r)$:
\begin{equation}
\left \{
\begin{array}{llll}
& \displaystyle - \, \left ( \frac{d}{dr} \,-\, \frac{\kappa_a}{r} \right )
F_a \,+\, V_a(r) \, G_a +m\, G_a &=& \displaystyle \varepsilon_a \, G_a
\\ \rule{0pt}{7mm} \displaystyle
 & \displaystyle  \, \left ( \frac{d}{dr} \,+\, \frac{\kappa_a}{r} \right )
G_a \,+\, V_a(r) \, F_a  \,-\,m \,F_a
&=& \displaystyle \varepsilon_a \, F_a \,.
\end{array} \right .
\label{dirac1}
\end{equation}
In contrast to the nonrelativistic Schr\"odinger equation,
generally speaking,
it is impossible to choose such a local potential  $V_a(r)$
which would exactly reproduce the one-electron energy $\veps_a$ and
 the radial components $G_a$ and $F_a$ for a given shell.
This is due to the fact that the potential $V_a(r)$ enters
both radial equations.  However, one can derive an approximate
potential by  inverting the radial Dirac equation for 
the large component: 
\begin{equation}
V^0_a(r) \,=\, \varepsilon_a \,-\,m\,
+ \,\frac{1}{G_a} \, \left ( \frac{d}{dr} \,-\, \frac{\kappa_a}{r} \right ) \, F_a
\,=\, V_{\rm C}+\frac{Y_a(r)}{r} \,+\, \frac{1}{G_a \,r}  \, X_a^F\,.
\end{equation}
This leads to a local potential   $V_a^0(r)$ which has some singularities,
because the function $G_a$ has nodes in the core region for $n_a > l_a+1$.

Let us consider another method of constructing the potential $V_a(r)$.
Multiplying the first and second radial Dirac equations with $G_a$ and
$F_a$, respectivily, and summing them,  we obtain
\begin{equation}
- \,  G_a \, \left ( \frac{d}{dr} - \frac{\kappa_a}{r} \right )
F_a + \, F_a \, \left ( \frac{d}{dr} + \frac{\kappa_a}{r} \right )G_a
\,+\, V_a(r) \, \rho_a \,+\,mG_a^2
- m F_a^2
\,=\, \displaystyle \varepsilon_a \, \rho_a \,,
\label{pot1}
\end{equation}
where $\rho_a \,=\, G_a^2+F_a^2$. Inverting this equation with respect
to $V_a(r)$, we have
\begin{equation}
\begin{array}{lll} \displaystyle V^{(1)}_a(r) &=& \displaystyle
\varepsilon_a \,+\, \frac{\,G_a}{\rho_a} \,
\left ( \frac{d}{dr}-\frac{\kappa_a}{r} \right ) F_a -
\frac{\, F_a}{\rho_a} \,
\left ( \frac{d}{dr} + \frac{\kappa_a}{r} \right ) G_a +
m\frac{F_a^2}{\rho_a}\,-\,m\frac{ G_a^2}{\rho_a}
\\ \rule{0pt}{9mm} &=& \displaystyle
V_{\rm C}+\frac{Y_a(r)}{r} \,+\, \frac{1}{\rho_a \,r} \,
\left ( G_a\, X_a^F \,+\, F_a\, X_a^G \right ) \,.
\end{array}
\end{equation}
Despite the potential  $V^{(1)}_a(r)$ has no singularities
in the core region, it can oscillate and singularities 
can occur in the nonrelativistic limit.

To smooth the potential $V^{(1)}_a(r)$ in the
core region, we use the following procedure. Instead of the density
$\rho_a$, we consider an average density 
$\overline \rho_a$ defined by
\begin{equation}
\overline \rho_{n_a \kappa_a} \,=\, \sum_{n \le n_a} \, w_{n\kappa_a} \,
\rho_{n \kappa_a} \,, \qquad \sum_{n \le n_a} \, w_{n\kappa_a} \,=\, 1 \,,
\end{equation}
where $w_{n\kappa_a}$ are positive weights. Thus, the density $\rho_{n_a \kappa_a}$
gets some admixture of the densities of the core shells corresponding to 
the same value of $\kappa_a$ but different values of the principal quantum
number $n < n_a$. Since  the maximal
values of the core shell densities are located nearby the nodes 
of the function $G_a$, the density 
$\overline \rho_{n_a \kappa_a}$ can be made to be  smooth and nodeless
 by a proper choice  of the weights
$w_{n\kappa_a}$. Outside the core region the densities
$\rho_{n_a \kappa_a}$  and $\overline \rho_{n_a \kappa_a}$
are almost coincide with each other. This is due to a fast decrease
of the core wave functions outside the core.
Assuming the nonlocal part of the DF potential 
can be replaced by a local potential which is the same
for all shells with the same $\kappa_a$, one can derive
\begin{equation}
V^{(2)}_a(r) \,=\, 
V_{\rm C}+\frac{Y_a(r)}{r} \,+\, \frac{1}{\overline \rho_{n_{a}\kappa_{a}} \,r} \, 
\sum_{n \le n_a} \,w_{n\kappa_a}
\left ( G_{n\kappa_a}\, X_{n\kappa_a}^F \,+\,
F_{n_a\kappa_a}\, X_{n \kappa_a}^G \right ) \,.
\end{equation}
 The potential 
 $V^{(2)}_a(r)$ derived for the shell $a$ can also be used
   for all shells with the same and different values of $\kappa_a$.
   This potential with weights
   $w_{n\kappa_a}\propto (m-\varepsilon_{n_a \kappa_a})/(m-\varepsilon_{n\kappa_a})$
   was used in our calculations.

In Table \ref{tab1}, we compare the energies of the cesium atom
obtained with the local potential
$V(r)$, that was derived using mainly the DF wave function of the $6s$ state,
with the DF energies and with the experimental ones. The corresponding
comparision for the francium atom, where the local potential
was  derived using mainly the DF wave function of the $7s$ state,
is presented in Table \ref{tab2}.

\subsection{Numerical evaluation of the QED corrections}

Numerical evaluation of expressions (\ref{zero})-(\ref{se_add})
was performed by employing the dual-kinetic-balance 
finite basis set method \cite{sha04} with basis
functions construced from B-splines. 
The calculation of
the zeroth-order contribution (\ref{zero}), 
with $V(r)$
constructed as indicated above, 
yields $E_{\rm PNC}=-$1.002 for $^{133}$Cs 
and $E_{\rm PNC}=-$10.19 for $^{223}$Fr,
in units of  i$\times 10^{-11}(-Q_W)/N$ a.u.
These values should be
compared with the corresponding DHF values, $-$0.741 for
$^{133}$Cs     and $-$13.72 
for $^{223}$Fr, and with the
 values that include the correlation effects, 
$-$0.904 for
$^{133}$Cs and $-$15.72 for  $^{223}$Fr
 (see the next section). 
The individual SE corrections
are presented in Table \ref{tab3}.
Since there is a significant cancellation between terms containing
the infrared singularities, the terms corresponding to
$n=a$ in  $\Sigma'_{\rm R}(\vare_a)$ and $n=b$ in  $\Sigma'_{\rm R}(\vare_a)$
are subtracted from contributions (\ref{se_a})-(\ref{se_d})  and added to 
 contributions  (\ref{se_k})-(\ref{se_l}).
The total SE correction
$\delta E_{\rm PNC}^{\rm tot}$, presented in Table  \ref{tab3},
 contains also the free term,
$-\alpha/(2\pi)E_{\rm PNC}$, mentioned above. Since this term is usually
included into the weak charge $Q_W$, one has to consider
the binding SE correction defined as 
$\delta E_{\rm PNC}^{\rm bind}=\delta E_{\rm PNC}^{\rm tot}
+\alpha/(2\pi)E_{\rm PNC}$. According to our calculations,
the binding SE correction amounts to $-$0.67\% for cesium
and $-$1.29\% for francium.
To estimate the uncertainty
of these values due to  correlation effects,
  we have also performed the calculations 
with $V(r)$ constructed employng the DF wave function of the
$7s$ state for cesium and the $8s$ state for francium.
 While this leads to a 
 2\% decrease of the transition amplitude,  the 
relative shift of the SE correction is, however, five times smaller.
Since the correlation effects contribute to the transition amplitude
on the 20\% level,
we assume a 4\% uncertainty for the total SE correction. 
Therefore, our value for the binding SE correction
is $-$0.67(3)\% for cesium and $-$1.29(5)\% for francium.
In case of cesium, 
our value differs 
from the previous evaluations of the SE effect, which are
$-$0.9(1)\% \cite{kuch03} and $-$0.85\% \cite{mil02}.

We have also calculated the VP correction.
The individual contributions for the Uehling part,
calculated including the screening correction as described after 
equation (\ref{uehlexpr}),
 are presented in Table \ref{tab4}.
The total Uehling correction is almost independent of the screening 
effect and amounts to 0.410\%  
for cesium and 1.037\% for francium.
These results agree
well with the previous calculations of this correction.
The individual contributions for the Wichmann-Kroll (WK) 
correction, obtained 
employing approximate formulas for the WK potential from
Ref. \cite{fain91}, are given in Table \ref{tab5}.
The total WK correction
is equal to  $-$0.004\% (cf. \cite{dzu02})
for cesium and $-$0.028\% for francium. 
This leads to the 0.406\%
result for the total VP correction for cesium and 
to the 1.01\%
result for francium.
Therefore, the total binding QED
correction amounts to $-$0.27(3)\%
for cesium and $-$0.28(5)\% for francium.

\section{Electron correlation effect on the PNC transition amplitude}

To calculate the correlation effects on
the PNC amplitude we start with the relativistic
 Hamiltonian in the no-pair approximation:
\begin{equation}
 H_{\rm np} = \Lambda_{+} H \Lambda_{+} \,,
\qquad
 H = \sum_j \, h_{\rm D}(j) + V_{\rm C}^{\rm int} +V_{\rm B}^{\rm int} \,,
\label{eq:Hnp}
\end{equation}
where $h_{\rm D}$ is the one-electron Dirac Hamiltonian, 
the index $j=1,...N$ enumerates the electrons, and 
$V_{\rm C}^{\rm int}$ 
and $V_{\rm B}^{\rm int}$ are
the Coulomb and the Breit electron-electron interaction operator,
respectively.
The frequency independent Breit interaction in the Coulomb gauge is given by
\begin{equation}
 V_{\rm B}^{\rm int} =  V_{\rm G}^{\rm int} \,+\, V_{\rm R}^{\rm int} \,, \qquad
 V_{\rm G}^{\rm int} =  - \,\alpha \sum_{i < j} \,
    \frac{{\balpha}_i \cdot {\balpha}_j}{r_{ij}} \,, \qquad
 V_{\rm R}^{\rm int} =  - \, \frac{\alpha}{2} \, \sum_{i < j} \,
            ({\balpha}_i \cdot {\bnabla}_i)
            ({\balpha}_j \cdot {\bnabla}_j)\, r_{ij} \,.
\end{equation}
Here $V_{\rm G}^{\rm int}$ is the so-called magnetic or Gaunt term and 
 $V_{\rm R}^{\rm int}$ is the
retardation term.
The operator $\Lambda_{+}$  is the projector
on the positive-energy states, which is the product of the one-electron
 projectors $\lambda_{+}(i)$,
\begin{equation}
\Lambda_{+} = \lambda_{+}(1) \cdot \cdot \cdot \lambda_{+}(N) \,,
\label{eq:Lambda}
\end{equation}
where
\begin{equation}
  \lambda_{+}(i) = \sum_{n} \mid u_n(i) \rangle \langle u_n(i) \mid \,.
\label{eq:lambda}
\end{equation}
Here $u_n(i)$ are the positive-energy eigenstates of an effective
one-particle Hamiltonian $h_u$,
\begin{equation}
h_u \, u_n \,=\, \varepsilon_n \, u_n \,,
\end{equation}
which can be taken to  be the Dirac Hamiltonian
$h_{\rm D}$, the Dirac Hamiltonian in an external field or 
the DF Hamiltonian in an external field
\cite{Sucher,Mittleman,gla04}.

To calculate the $E_{\rm PNC}$ amplitude, we add the weak
interaction to the full Hamiltonian:
\begin{equation}
H(\mu) \,=\, H \,+\, \mu \, \sum_j \, H_{W}(j)\,,
\end{equation}
where $H_W$ is defined by Eq. (\ref{h_w}).

With the PNC interaction added to the one-electron DF Hamiltonian,
one obtains the coupled equations, which are usually referred to as
the PNC-HF equations \cite{Sandars1}. The linearization of these equations
with respect  to the
parameter $\mu$ would make them inhomogeneous. Since
in our calculations we do not
perform such a linearization, 
the  equations remain homogeneous. In this case the PNC amplitude
can be calculated using the equation
\begin{equation}
  E_{\rm PNC} = \frac{\partial}{\partial \mu} \Bigl [
    \langle \Psi^{f} (\mu) \mid D_{z} \mid \Psi^{i} (\mu) \rangle
  \Bigr ]_{\mu=0} \,,
\label{e_pnc}
\end{equation}
where ${\bf D} = \sum_i e{\bf r}_i $ is the dipole moment operator
 and
$\Psi^i$ and $\Psi^f$ are the many-electron wave functions of the initial and
final states, respectively.  They obey the equations
\begin{equation}
H(\mu) \, \Psi^i(\mu) \,=\, E_i(\mu) \, \Psi^i(\mu) \,, \qquad
H(\mu) \, \Psi^f(\mu) \,=\, E_f(\mu) \, \Psi^f(\mu) \,.
\end{equation}
The many-electron wave functions $\Psi^i$ and $\Psi^f$ are represented by
a large number of the configuration state functions (CSFs):
\begin{equation}
\Psi^{J M}(\mu) \,=\, \sum_{\alpha} C_{\alpha}(\mu) \,
\Phi_{\alpha}^{J M}(\mu) \,.
\label{eq:Psi}
\end{equation}
The CSFs $\Phi_{\alpha}^{J M}$ are  linear combinations of the
Slater determinants, which are constructed from the one-electron wave
functions $u_n(\mu)$.
Expansion (\ref{eq:Psi}) contains the CFSs of different parity,
since the weak interaction is included in the Hamiltonian $H(\mu)$.

The one-electron functions $u_n(\mu)$ are obtained as 
eigenfunctions of the Dirac-Fock operator in the external field:
\begin{equation}
h_u(\mu) \, u_n(\mu) \,=\, \varepsilon_n(\mu) \, u_n(\mu) \,, \qquad
h_u(\mu) \,=\, h_{\rm DF}(\mu) \,+\, \mu \,H_{W}\,.
\label{DF1} 
\end{equation}
It should be noted that the Dirac-Fock operator $h_{\rm DF}(\mu)$ depends
on the parameter $\mu$, since the one-particle density matrix is
constructed from occupied orbitals $u_n(\mu)$.
We can also consider the set of one-electron wave functions $u^0_n(\mu)$
defined by equations
\begin{equation}
h^0_u(\mu) \, u^0_n(\mu) \,=\, \varepsilon^0_n(\mu) \, u^0_n(\mu) \,, \qquad
h^0_u(\mu) \,=\, h_{\rm DF}(0) \,+\, \mu \,H_{W}\,,
\label{DF2}
\end{equation}
where $h_{\rm DF}(0)$ is the standard Dirac-Fock operator without
the external field.

The PNC amplitude can be calculated in the Hartree-Fock approximation by using
only one CSF  in expansion (\ref{eq:Psi}). Using 
equation (\ref{e_pnc}) and the wave functions $u^0_n(\mu)$,
one obtains so-called Dirac-Fock value of the PNC amplitude.
If the set of $u_n(\mu)$ is used, the method, in principle, is equivalent
to the PNC-HF method, which was used by different authors
\cite{Sandars2,Johnson_85,Johnson_86}.

 In the large scale configuration interaction (CI) method the set of
the CSFs for given quantum numbers $JM$ is generated including all
single, double, and the most significant part of triple excitations in
the positive spectrum of the one-electron states $u_n(\mu)$.
In what follows, this method of evaluation of
 the PNC amplitude will be
referred to as the PNC-CI method.

To obtain the set of the one-electron functions $u_n(\mu)$ and $u^0_n(\mu)$,
we solve  equations (\ref{DF1}) and (\ref{DF2}) using the finite
basis approximation,
\begin{equation}
u_n(\mu) \,=\, \sum_{a} \, c^n_{a}(\mu) \, \varphi_a \, 
\end{equation}
with the basis functions $\varphi_a$ given in the central field
approximation:
\begin{equation}
\varphi_{a}({\bf r}) \,=\,
\frac{i^{l_a}}{r}
\left  ( \begin{array}{cc}
  P_a(r) &  \chi_{\kappa_a m_a}({\bf n}) \\
i Q_a(r) & \chi_{-\kappa_a m_a}({\bf n})
\end{array} \right )\,.
\label{basis}
\end{equation}
The representation (\ref{basis}) differs from the standard one
by the factor $i^l$. This factor is introduced
to make the one-electron matrix elements of the PNC Hamiltonian
to be real:
\begin{equation} \label{pnc_rad}
\langle a \mid  \gamma^5 \rho_N \mid b \rangle \,=\,
-\, (-1)^{(l_b-l_a+1)/2} \, \delta_{\kappa_a,-\kappa_b} \,
\delta_{m_a, m_b} \,
\int \limits_0^{\infty} \, dr\; \rho_N
\left [ P_a \, Q_b \,-\, Q_a \, P_b \right]\,,
\end{equation}
where $l_a+l_b+1$ is even. With this one-electron basis, the
large scale PNC-CI matrix is also real and Hermitian.

For the occupied atomic shells, the large $P_a $ and small $Q_a$ components
of the radial wave functions are obtained by solving the standard 
radial DF equations. For the vacant shells the Dirac-Fock-Sturm
equations are used. For details of the Dirac-Fock-Sturm method
we refer to Refs. \cite{tup_03,gla04}.
The basis set containing  the radial functions up to 
17s, 16p, 12d, 7f, 5g, and 2h states was used in the calculations.

In the calculations of the one-electron PNC matrix elements (\ref{pnc_rad})
we used the Fermi nuclear distribution
\begin{equation}
\rho_N(r) \,=\, \frac{\rho_0}{1+e^{4\,{\rm ln}3 \,(r-c)/t}}\,,
\end{equation}
where $t=2.3$ fm. The parameters $c$ and $\rho_0$ were determined to reproduce
the value of the nuclear mean-square radius $R_N=\la r^2\ra^{1/2}$ and the
normalization condition for $\rho_N(r)$.

In Table \ref{tab6} we present the results of our calculations of the
PNC amplitude for Rb, Cs, and Fr. The results obtained by the DF method
are given in the third column.
 Our DF value for the
$6s$-$7s$ PNC transition in Cs,  $-$0.741,
is in a good agreement with the values
$-$0.742 \cite{koz01} and $-$0.739 \cite{dzu01}, which were obtained
by the direct summation over the intermediate states. For the 
$7s$-$8s$ PNC transition in Fr our DF value, $-$13.72, is also in a good
agreement with the $-$13.56 result obtained in Ref. \cite{saf00}.
Our PNC-HF values, $-$0.138 for Rb and $-$0.926 for Cs,
can be compared with the values $-$0.139 and $-$0.927,
respectively, obtained by a similar method in Ref. \cite{Johnson_86}.
In the fifth column of the table, we present our PNC-CI values, which include
the core-polarization correlation effects. 
The uncertainty of these values is estimated to be on the
1\% level.
For comparison, the most accurate results by other authors are
listed in the sixth column of the table.
In the second column we give the values of the 
nuclear-mean-square radius $R_N$, which were used in our calculations.
They were obtained by the formula $R_N=0.836A^{1/3}+0.570$ \cite{joh85}.
In case of Fr, the corresponding results with $R_N$ taken from
Ref. \cite{ang04} are also presented.

To calculate the contribution of the  frequency independent Breit interaction
(BI) to the PNC amplitude, we included the magnetic $V^{\rm int}_{\rm G}$ and
retardation $V^{\rm int}_{\rm R}$ terms in all stages of the calculations. 
As the first
step, the BI was included in the radial Dirac-Fock equations. We will refer this
approach to as the Dirac-Fock-Breit (DFB) method. On the second stage, the BI
was added to  the Dirac-Fock-Sturm equations and to the Dirac-Fock Hamiltonian
$h_u(\mu)$ in the external field (\ref{DF1}). This method of calculation of the
PNC amplitude  will be called as the PNC-HFB method.
Finally, we added the BI to
the many-electron Hamiltonian $H(\mu)$ in the external field and performed
the large scale CI calculation. This approach will be called as the PNC-CIB 
method.
To estimate the role of the retardation part of the Breit intaraction,
we  repeated all the calculations including only the
magnetic (Gaunt) part $V^{\rm int}_{\rm G}$ of the BI and then took the
difference with the PNC amplitude, which includes the total BI.

In Table \ref{tab7} we present the magnetic Breit $\delta E^{\rm M}_{\rm PNC}$
and retardation Breit $\delta E^{\rm R}_{\rm PNC}$ contibutions to 
the $6s$-$7s$ PNC amplitude in $^{133}$Cs and to 
the $7s$-$8s$ PNC amplitude in $^{223}$Fr,
obtained by different methods. 
The comparison
of the total Breit correction to the PNC amplitude with the most accurate
results by other authors are presented in Table \ref{tab8}.
Finally, in case of francium, our PNC-CIB value amounts to 
$-$15.58(16) [$R_N = 5.640$fm] and  $-$15.55(16) [$R_N = 5.658$fm]  
for $^{223}$Fr, and $-$14.21(14) for $^{210}$Fr.
They are in a fair agreement with the most accurate previous results
\cite{saf00},
$-$15.41(17) [$R_N = 5.640$fm]
for $^{223}$Fr and $-$14.02(15) for $^{210}$Fr.

\section{Total PNC amplitudes}
To get the total $6s$-$7s$ PNC transition amplitude in $^{133}$Cs,
we combine the most accurate  value that includes the correlation
and Breit effects \cite{dzu02}, $-$0.902(5),
with the $-$0.27(3)\% binding QED correction, the $-$0.19(6)\%
neutron skin correction \cite{der01}, the $-$0.08\%
correction due to the renormalization of $Q_W$ from the
atomic momentum transfer $q\sim 30$ MeV down to $q=0$
\cite{mil02}, and the 0.04\% contribution from the 
electron-electron weak interaction \cite{mil02,sus78}. 
The analysis of accuracy of the atomic structure PNC calculations 
\cite{koz01,dzu02,dzu89,blu00}
 is based on calculations of the hyperfine splitting, decay rates, and
 energy levels. As it was argued in Ref. \cite{mil02}, QED corrections to these
 quantities can be neglected on the 0.5\% accuracy level.
Using the experimental
value for  $E_{\rm PNC}/\beta$
\cite{wood97} and an average value from two most accurate measurements of the
vector transition polarizabilty, $\beta=26.99(5)a_{\rm B}^3$
  \cite{ben99,dzu02,cho97,vas02}, 
we obtain for the weak charge
of $^{133}$Cs:
\begin{eqnarray}
Q_W=-72.65(29)_{\rm exp}(36)_{\rm th}\,.
\end{eqnarray}
This value
deviates from the SM prediction
 of $-$73.19(13) \cite{ros02}
 by 1.1 $\sigma$.

In case of francium, combining our PNC-CIB values,
$-$15.55(16) for $^{223}$Fr and $-$14.21(14) for $^{210}$Fr,
with the $-$0.28(5)\% QED correction 
and the $-$0.08\%
correction due to the renormalization of $Q_W$ from the
atomic momentum transfer $q\sim 30$ MeV down to $q=0$
\cite{mil02}, we obtain 
$-$15.49(16) for $^{223}$Fr and $-$14.16(14) for $^{210}$Fr.

In summary, we have calculated the QED correction to the
 PNC transition amplitude in Cs and Fr. 
In addition, we have performed
 an independent high-precision calculation of the correlation
and Breit interaction effects on the PNC amplitude in Fr. 
We have derived
the weak charge of $^{133}$Cs, which deviates by 1.1$\sigma$
from the SM prediction.
Further improvement of atomic tests of the standard model can be achieved,
from theoretical side,  by more accurate calculations
of the electron-correlation effects and, from experimental side, by more
precise measurements of the PNC amplitude in cesium or other atomic systems,
particularly, in francium \cite{beh93,cal05}.

\section*{Acknowledgements}

Valuable discussions with K.T. Cheng, V.A. Dzuba, V.V. Flambaum,
M.Y. Kuchiev, M.S. Safronova,
and O.P. Sushkov are gratefully acknowledged.
This work was supported by EU (Grant No. HPRI-CT-2001-50034),
RFBR (Grant No. 04-02-17574), 
NATO (Grant No. PST.CLG.979624), and DFG.  

\appendix

\section{Zero-potential PNC vertex contribution}

The zero-potential PNC vertex contribution is defined as
\be \label{lam_w}
\la b|\Lambda_W|a\ra \equiv \int \frac{d {\bf p}}{(2\pi)^3} \int
\frac{d \bfp'}{(2\pi)^3}\;
\overline{\psi}_b(\bfp')\Gamma_W^0(\rrp',\rrp)
V_W(|\bfp'-\bfp|)
\psi_a(\bfp)\,,
\ee
where $\rrp=(\veps,\bfp)$ and $\rrp'=(\veps',\bfp')$ are four vectors,  
\be
\Gamma^0_W(\rrp',\rrp)  =-4\pi i\alpha\int \frac{d^4 \rrk}{(2\pi)^4}\gamma_{\sigma}
\frac{\cross{\rrp}'-\cross{\rrk}+m}{(\rrp'-\rrk)^2-m^2}\gamma^0\gamma^5 
\frac{\cross{\rrp}-\cross{\rrk}+m}{(\rrp-\rrk)^2-m^2}\gamma^{\sigma}\frac{1}{\rrk^2}\,,
\ee
\be
V_W(q) \equiv \eta\int d\bfr \rho_N(r)\exp{(i\bfq\cdot \bfr)}
=\eta\frac{4\pi}{q}\int_0^{\infty} d r \, r\rho_N(r)\sin{(qr)}\,,
\ee
$q=|\bfq|$, $r=|\bfr|$,  and $\eta = -(G_F/\sqrt{8})Q_W$.
In equation (\ref{lam_w}), it is implicit that $\veps=\veps_a$
and $\veps'=\veps_b$.
One can easily express $\Gamma^0_W(p',p)$ in terms of the standard 
vertex function $\Gamma^0(p',p)$:
\be
\Gamma^0_W(\rrp',\rrp)=\Gamma^0(\rrp',\rrp)\gamma^5 -\frac{\alpha}{\pi}
[2\veps'm(C_0+C_{11})+2\veps m C_{12}-m^2\gamma^0 C_0]\gamma^5\,,
\ee
where 
the coefficients 
$C_0$, $C_{11}$, and $C_{12}$ are defined as in Ref. \cite{yer99}.
After the isolation of the ultraviolet divergences in
$\Gamma^0(\rrp',\rrp)$,
the finite part is given by
\be
\Gamma^0_{W,R}(\rrp',\rrp)&=&\frac{\alpha}{4\pi}
[(A+4m^2C_0)\gamma_0 + \cross{\rrp}'(B_1\veps' +B_2\veps)
+\cross{\rrp}(C_1\veps'+C_2\veps)+D \cross{\rrp}'\gamma_0\cross{\rrp}\nonumber\\
&&+H_1\veps'+H_2\veps -8\veps'm(C_0+C_{11})
-8\veps m C_{12}]\gamma^5\,,
\ee
where
all the coefficients  are defined as in Ref. \cite{yer99}.
Integrating over  the angles in Eq. (\ref{lam_w}), one can obtain
\be \label{lam_w_f}
\la b|\Lambda_{W,R}|a\ra &=& -\frac{\alpha}{2(2\pi)^6}i^{l_b-l_a}
\delta_{\kappa_b,-\kappa_a} \delta_{m_b,m_a}
\int_0^{\infty} dp \int_0^{\infty} dp' p^2 p'^2 \int_{-1}^{1} d\xi \;
[V_W(q) Q_1(p',p,\xi) P_{l_b}(\xi)\nonumber\\
&& + V_W(q) Q_2(p',p,\xi) P_{l_a}(\xi)]\,,
\ee
where $P_l(\xi)$ is a Legendre polynomial, $\kappa = (-1)^{j+l+1/2}(j+1/2)$,
\be
Q_1&=&[A+4m^2C_0+\veps'(B_1\veps'+B_2\veps) +\veps(C_1\veps'+C_2\veps)
+D\veps'\veps\nonumber\\
&&+ H_1\veps' +H_2\veps -8\veps'm(C_0+C_{11}) -8\veps m C_{12}]
\tilde{g}_b(p') \tilde{f}_a(p)\nonumber\\
&& + [p'(B_1\veps'+B_2\veps) + Dp'\veps] \tilde{f}_b(p') \tilde{f}_a(p)
\nonumber\\
&&+[p(C_1\veps'+C_2\veps) + Dp\veps'] \tilde{g}_b(p') \tilde{g}_a(p)
+Dp'p \tilde{f}_b(p') \tilde{g}_a(p)\,,
\ee
\be
Q_2&=&[A+4m^2C_0+\veps'(B_1\veps'+B_2\veps) +\veps(C_1\veps'+C_2\veps)
+D\veps'\veps\nonumber\\
&&- H_1\veps' -H_2\veps +8\veps'm(C_0+C_{11}) +8\veps m C_{12}]
\tilde{f}_b(p') \tilde{g}_a(p)\nonumber\\
&& + [p'(B_1\veps'+B_2\veps) + Dp'\veps] \tilde{g}_b(p') \tilde{g}_a(p)
\nonumber\\
&&+[p(C_1\veps'+C_2\veps) + Dp\veps'] \tilde{f}_b(p') \tilde{f}_a(p)
+Dp'p \tilde{g}_b(p') \tilde{f}_a(p)\,,
\ee
$\tilde{g}(p)$ and $\tilde{f}(p)$ are the radial components of the Dirac
wave function in the momentum representation, 
defined as in Ref. \cite{yer99}.

\newpage

\begin{table}
\caption{The binding energies of low-lying states in Cs, in a.u.
The experimental energies are taken from Ref. \cite{cs} } \vspace{0.2cm}
\label{tab1}
\begin{ruledtabular}
\begin{tabular} {cccc}
State &\multicolumn{1}{c}{Local potential}
                 & \multicolumn{1}{c}{DF}
                                     & \multicolumn{1}{c}{Exp.} \\
\hline
$6s_{1/2}$ & -0.13079 & -0.12824  & -0.14310 \hspace*{0.115cm}
\\
$6p_{1/2}$ & -0.08696 &  -0.08582 & -0.09217 \hspace*{0.115cm}
\\
$6p_{3/2}$ & -0.08479 &  -0.08397 & -0.08965 \hspace*{0.115cm}
 \\
$7s_{1/2}$ & -0.05621   & -0.05537 & -0.05865 \hspace*{0.115cm}
\\
$7p_{1/2}$ & -0.04251 &  -0.04209 & -0.04393 \hspace*{0.115cm}
\\
$7p_{3/2}$ & -0.04175 &  -0.04143 &  -0.04310 \hspace*{0.115cm}
\end{tabular}
\end{ruledtabular}
\end{table}

\begin{table}
\caption{The binding energies of low-lying states in Fr, in a.u.} \vspace{0.2cm}
\label{tab2}
\begin{ruledtabular}
\begin{tabular} {cccc}
State &\multicolumn{1}{c}{Local potential}
                 & \multicolumn{1}{c}{DF}
                                     & \multicolumn{1}{c}{Exp. \cite{fr1,fr2,fr3,fr4,fr5,fr6,fr7}} \\
\hline
$7s_{1/2}$ & -0.13640 & -0.13271  & -0.14967 \hspace*{0.115cm}
\\
$7p_{1/2}$ & -0.08857 &  -0.08629 & -0.09391 \hspace*{0.115cm}
\\
$7p_{3/2}$ & -0.08199 &  -0.08071 & -0.08623 \hspace*{0.115cm}
 \\
$8s_{1/2}$ & -0.05740   & -0.05626 & -0.05976 \hspace*{0.115cm}
\\
$8p_{1/2}$ & -0.04297 &  -0.04222 & -0.04436 \hspace*{0.115cm}
\\
$8p_{3/2}$ & -0.04071 &  -0.04023 &  -0.04188 \hspace*{0.115cm}
\end{tabular}
\end{ruledtabular}
\end{table}

\begin{table}
\caption{The SE corrections to the $6s$-$7s$ PNC amplitude in Cs
and to the $7s$-$8s$ PNC amplitude in Fr, 
in \%. The results are presented in both the length (L)
and the velocity (V) gauge.} \vspace{0.2cm} 
\label{tab3}
\begin{ruledtabular}
\begin{tabular} {ccccc}
  &\multicolumn{2}{c}{Cs}
                 & \multicolumn{2}{c}{Fr} \\
\hline
Contribution & L-gauge  & V-gauge & L-gauge  & V-gauge 
\hspace*{0.115cm}
\\
\hline
$\delta E_{\rm PNC}^{\rm a}$ &-0.09 & -0.11 
 &0.18 & 0.15 \hspace*{0.115cm}
\\
$\delta E_{\rm PNC}^{\rm b}$ & 1.31 & 1.11 
& 1.84 & 1.35\hspace*{0.115cm}
\\
$\delta E_{\rm PNC}^{\rm c}$ & 0.34 & 0.40 
& -0.36 & -0.23\hspace*{0.115cm}
\\
$\delta E_{\rm PNC}^{\rm d}$ & -0.38 & -0.32 
 & -0.64 & -0.51\hspace*{0.115cm}
\\
$\delta E_{\rm PNC}^{\rm e}$ &-1.29 & -1.53  
 &-1.21 & -1.46 \hspace*{0.115cm}
\\
$\delta E_{\rm PNC}^{\rm f}$ & 3.89 & 3.25  
 & 3.61 & 2.94 \hspace*{0.115cm}
\\
$\delta E_{\rm PNC}^{\rm g}$ & 1.33 & 1.57  
& 1.32 & 1.58\hspace*{0.115cm}
\\
$\delta E_{\rm PNC}^{\rm h}$ & -4.04 & -3.40 
 & -4.03 & -3.36 \hspace*{0.115cm}
\\
$\delta E_{\rm PNC}^{\rm i}$ & -4.61 & -3.97 & -4.97 & -4.30
\hspace*{0.115cm}
\\
$\delta E_{\rm PNC}^{\rm j}$ & 1.49 & 1.73
& 1.58 & 1.83
 \hspace*{0.115cm}
\\
$\delta E_{\rm PNC}^{\rm k}$ & -0.79 & -1.03 
 & -0.78 & -1.04
\hspace*{0.115cm}
\\
$\delta E_{\rm PNC}^{\rm l}$ & 2.05 & 1.41 
& 2.05 & 1.38
\hspace*{0.115cm}
\\
$\delta E_{\rm PNC}^{\rm add}$ & 0.00 & 0.10 & 0.00 & 0.26
\hspace*{0.115cm}
\\
$\delta E_{\rm PNC}^{\rm tot}$ & -0.79 & -0.79 & -1.40 & -1.40
\hspace*{0.115cm}
\end{tabular}
\end{ruledtabular}
\end{table}

\begin{table}
\caption{The Uehling corrections to the $6s$-$7s$ PNC amplitude in Cs
and  to the $7s$-$8s$ PNC amplitude in Fr,
in \%. The results are presented in both the length (L)
and the velocity (V) gauge.} 
\label{tab4}
\vspace{0.2cm} 
\begin{ruledtabular}
\begin{tabular} {ccccc}
  &\multicolumn{2}{c}{Cs}
                 & \multicolumn{2}{c}{Fr} \\
\hline
Contribution & L-gauge  & V-gauge & L-gauge  & V-gauge 
\hspace*{0.115cm}
\\
\hline
$\delta E_{\rm PNC}^{\rm a}$ 
 & -0.026  &-0.024 & -0.107 &-0.100 \hspace*{0.115cm}
\\
$\delta E_{\rm PNC}^{\rm b}$ 
 &- 0.050  &-0.024 & -0.208 &-0.098 \hspace*{0.115cm}
\\
$\delta E_{\rm PNC}^{\rm c}$ 
 & 0.354  &0.347 & 0.930 &0.902 \hspace*{0.115cm}
\\
$\delta E_{\rm PNC}^{\rm d}$ 
 & -0.054  &-0.061 & -0.077 &-0.107 \hspace*{0.115cm}
\\
$\delta E_{\rm PNC}^{\rm e}$ 
 & -0.070  &-0.069 & -0.188 &-0.188 \hspace*{0.115cm}
\\
$\delta E_{\rm PNC}^{\rm f}$ 
 & 0.255  &0.256 & 0.687 &0.687 \hspace*{0.115cm}
\\
$\delta E_{\rm PNC}^{\rm add}$ 
 & 0  &-0.014 & 0 &-0.060 \hspace*{0.115cm}
\\
$\delta E_{\rm PNC}^{\rm tot}$ 
 & 0.410  &0.410 & 1.037 &1.037 \hspace*{0.115cm}
\end{tabular}
\end{ruledtabular}
\end{table}

\begin{table}
\caption{The Wichmann-Kroll corrections to the $6s$-$7s$ PNC amplitude in Cs
and  to the $7s$-$8s$ PNC amplitude in Fr,
in \%. The results are presented in both the length (L)
and the velocity (V) gauge.} 
\label{tab5}
\vspace{0.2cm} 
\begin{ruledtabular}
\begin{tabular} {ccccc}
  &\multicolumn{2}{c}{Cs}
                 & \multicolumn{2}{c}{Fr} \\
\hline
Contribution & L-gauge  & V-gauge & L-gauge  & V-gauge 
\hspace*{0.115cm}
\\
\hline
$\delta E_{\rm PNC}^{\rm a}$ 
 & 0.0006  &0.0006 & 0.0053 &0.0049 \hspace*{0.115cm}
\\
$\delta E_{\rm PNC}^{\rm b}$ 
 & 0.0012  &0.0006 & 0.0102 &0.0048 \hspace*{0.115cm}
\\
$\delta E_{\rm PNC}^{\rm c}$ 
 & -0.0042  &-0.0041 & -0.0284 &-0.0270 \hspace*{0.115cm}
\\
$\delta E_{\rm PNC}^{\rm d}$ 
 & 0.0001  &0.0003 & -0.0009 &0.0006 \hspace*{0.115cm}
\\
$\delta E_{\rm PNC}^{\rm e}$ 
 & 0.0007  &0.0007 & 0.0055 &0.0055 \hspace*{0.115cm}
\\
$\delta E_{\rm PNC}^{\rm f}$ 
 & -0.0026  &-0.0026 & -0.0199 &-0.0199 \hspace*{0.115cm}
\\
$\delta E_{\rm PNC}^{\rm add}$ 
 & 0  &0.0003 & 0 &0.0030 \hspace*{0.115cm}
\\
$\delta E_{\rm PNC}^{\rm tot}$ 
 & -0.0042  &-0.0042 & -0.0283 &-0.0283 \hspace*{0.115cm}
\end{tabular}
\end{ruledtabular}
\end{table}


\begin{table}
\caption{The $E_{\rm PNC}$ amplitude, 
in units of  i$\times 10^{-11}(-Q_W)/N$ a.u.,
calculated by different methods without the Breit correction.}
\label{tab6}
\vspace{0.2cm}
\begin{ruledtabular}
\begin{tabular} {clrrrr}
    &  \multicolumn{1}{c}{$R_N$[fm]}
    &\multicolumn{1}{c}{DF}
                 & \multicolumn{1}{c}{PNC-HF}
                                     & \multicolumn{1}{c}{PNC-CI}
                                     & \multicolumn{1}{c}{Others} \\
\hline
$ ^{85}$Rb \hspace{4mm} 5s $\to$ 6s & 4.246 & -0.110 & -0.138 & -0.134 & -0.135$^a$
\hspace*{0.115cm}
\\
$^{133}$Cs \hspace{4mm} 6s $\to$ 7s & 4.837 & -0.741 & -0.926 & -0.904 & -0.906$^b$
\hspace*{0.115cm} \\[-2mm]
                                    & &       &       &       & -0.908$^c$
\hspace*{0.115cm} \\
$^{223}$Fr \hspace{4mm} 7s $\to$ 8s & 5.640  & -13.72 & -16.63 & -15.72 & -15.56$^d$
\hspace*{0.115cm} \\[-2mm]
                                    & &       &       &       & -15.8$^e\,\,$
\hspace*{0.115cm} \\[-2mm]
                                    & 5.658 \cite{ang04} &  -13.69    &  -16.60     &  -15.69    & 

\hspace*{0.115cm} \\
$^{210}$Fr \hspace{4mm} 7s $\to$ 8s & 5.539 & -12.51 & -15.17 &  -14.34   &  
\hspace*{0.115cm} \\[-2mm]
                                    &5.545 \cite{ang04}&  -12.51     &  -15.16      &  -14.34    & 
\hspace*{0.115cm}
\end{tabular}
\end{ruledtabular}
%
\begin{flushleft}
\vspace{-2mm}
\noindent  $^a$ PNC-HF+MBPT \cite{Johnson_86}. \\[-2mm]
\noindent  $^b$ MBPT \cite{Johnson_90}.\\[-2mm]
\noindent  $^c$ Correlation Potential+MBPT \cite{dzu02}.\\[-2mm]
\noindent  $^d$ MBPT \cite{saf00}.\\[-2mm]
\noindent  $^e$ Correlation Potential+MBPT \cite{dzu95}. The original
value, $-$15.9 \cite{dzu95}, is rescaled to $R_N=5.640$ according to
the corresponding analysis presented in Ref. \cite{saf00}.
\end{flushleft}
\end{table}

%
\begin{table}
\caption{The Breit magnetic ($\delta E^{\rm M}_{\rm PNC}$), the Breit retardation
($\delta E^{\rm R}_{\rm PNC}$), and the total Breit ($\delta E^{\rm B}_{\rm PNC}$) 
correction
to the PNC amplitude, in units of 
 i$\times 10^{-11}(-Q_W)/N$ a.u.
}
\label{tab7}
\vspace{0.2cm}
\begin{ruledtabular}
\begin{tabular} {ccrrr}
    &    &\multicolumn{1}{c}{DFB}
                 & \multicolumn{1}{c}{PNC-HFB}
                                     & \multicolumn{1}{c}{PNC-CIB} \\
\hline
$^{133}$Cs & $\delta E^{\rm M}_{\rm PNC}$  & 0.0028   & 0.0023 & 0.0049 \\[-1.5mm]
           & $\delta E^{\rm R}_{\rm PNC}$  & -0.0006   & -0.0005 & -0.0004 \\[-1.5mm]
           & $\delta E^{\rm B}_{\rm PNC}$  & 0.0022   & 0.0018 & 0.0045 \\[1.5mm]
$^{223}$Fr &  $\delta E^{\rm M}_{\rm PNC}$ & 0.080    & 0.082  & 0.165 \\[-1.5mm]
           &  $\delta E^{\rm R}_{\rm PNC}$ & -0.016   &  -0.017  & -0.022 \\[-1.5mm]
           &  $\delta E^{\rm B}_{\rm PNC}$ & 0.064    & 0.065  & 0.143
\end{tabular}
\end{ruledtabular}
%
\end{table}
\pagebreak
\begin{table}
\caption{ Comparison of the total Breit correction to the PNC amplitude,
in units of i$\times 10^{-11}(-Q_W)/N$ a.u.,
with the most accurate results by
other authors.} 
\label{tab8}
\begin{ruledtabular}
\begin{tabular}{lllll}
\multicolumn{2}{c}{$^{133}$Cs}
&\multicolumn{1}{c}{}
& \multicolumn{2}{c}{$^{223}$Fr} \\
\hline
This work                        & 0.0045 && This work   & 0.14
\\[-1mm]
Kozlov {\it et al.}\cite{koz01} & 0.004  && Safronova and Johnson
\cite{saf00} & 0.15 \\[-1mm]
Dzuba {\it et al.}\cite{dzu02} & 0.0055 && Derevianko \cite{der00a} & 0.18
\\[-1mm]
Derevianko \cite{der01} & 0.0054 &&    &
\\
\end{tabular}
\end{ruledtabular}
\end{table}


\begin{widetext}
\begin{figure}
\begin{minipage}{16cm}
\centering
\includegraphics[clip=true,width=0.5\textwidth]{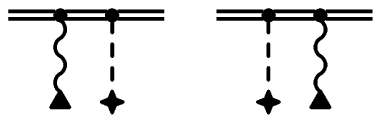}
\caption{Feynman diagrams for the lowest-order PNC transition
amplitude. The wavy line terminated
with a triangle indicates the absorbed photon. The dashed 
line terminated with a cross indicates the electron-nucleus weak interaction.
 \label{fig:pnc}}
\end{minipage}
\end{figure}

\begin{figure}
\begin{minipage}{16cm}
\centering
\includegraphics[clip=true,width=1.0\textwidth]{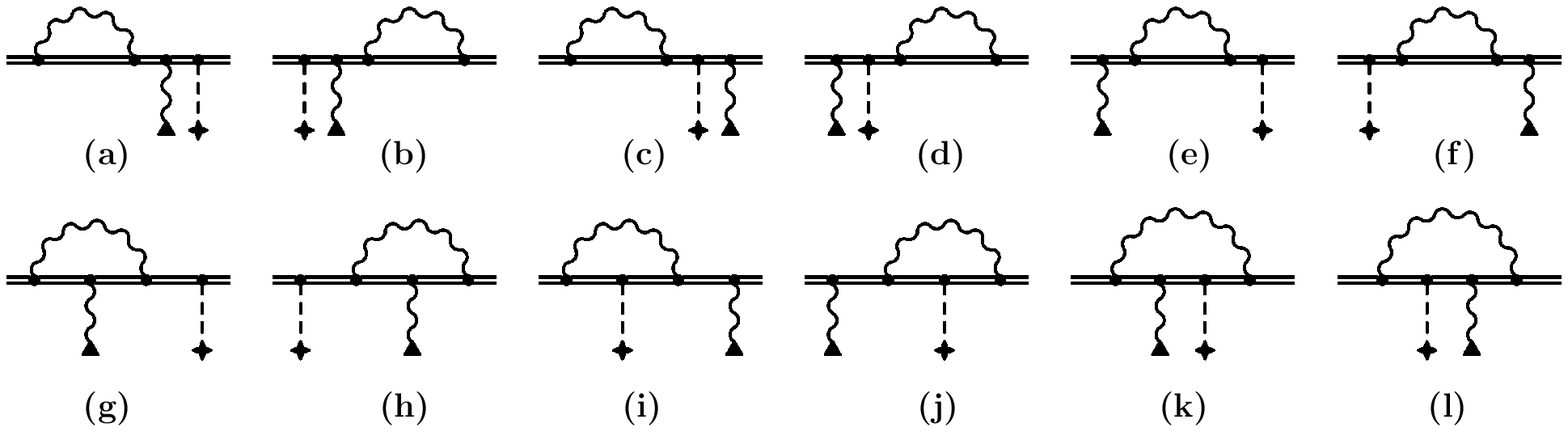}
\caption{Feynman diagrams for the SE corrections to
the PNC transition amplitude. The wavy line terminated
with a triangle indicates the absorbed photon. The dashed 
line terminated with a cross indicates the electron-nucleus weak interaction.
 \label{fig:sepnc}}
\end{minipage}
\end{figure}
\end{widetext}


\end{document}